\documentclass[twocolumn,secnumarabic,showpacs,preprintnumbers,amssymb,floatfix,prb]{revtex4}
\usepackage{amsmath,amssymb}
\usepackage{graphicx}

\begin{document}

\title{Photogalvanic effect and photoconductance
in a quantum channel with a single
short-range scatterer}

\author{M. A. Pyataev}

\email{pyataevma@math.mrsu.ru}

\author{S. N. Ulyanov}

\affiliation{Mordovian State University, Saransk, Russia}.

\date{\today}

\begin{abstract}
The influence of electromagnetic radiation on
the electron transport in a quantum channel
with a single short-range scatterer is investigated
using a generalized Landauer-B\"{u}ttiker approach.
We have shown that asymmetrical position of the scatterer
leads to appearance of the direct photocurrent in the system.
The dependence of the photocurrent on  the electron chemical
potential, the position of the scatterer, and the frequency of the radiation is studied.
We have shown that the photocurrent and the photoconductance oscillate
as functions of the electron chemical potential.
The nature of oscillations is discussed.
\end{abstract}

\pacs{73.23.Ad, 78.67.Lt}

\maketitle

\section{Introduction}

The study of electron transport in nanostructures
under the external electromagnetic radiation
has been attracting considerable attention
during the last few years. The interest to the problem is
stipulated by recent attempts to create a sensitive
nanometer-size photodetector. \cite{Kos02,Kos03,Rao05,Row06,Ejr07}
Photon-induced electron transport in a number of interesting systems
such as quantum point contacts, \cite{Wyss93,Wyss95,Wir02,Kos02}
field-effect transistors, \cite{Kos03,Row06}
quantum dots, \cite{Rao05}
and carbon nanotubes \cite{Mul04,Cho07}
was studied in various experiments.
Quantum ratchet effects induced by terahertz radiation
were observed in GaN-based two-dimensional structures. \cite{Web08}
Impurity photocurrent spectra of bulk GaAs and GaAs quantum wells
doped with shallow donors were studied. \cite{Ale07}
It was shown experimentally \cite{Wyss93}
that the photoconductance of the quantum point contact
oscillates with the gate voltage.
In the case of asymmetrical illumination,
the direct photocurrent arises in the quantum point contact. \cite{Wyss95}.
The photocurrent was attributed to the radiation-induced thermopower.

A lot of theoretical models have been used to study the
microwave induced electron transport
in bulk materials and low-dimensional systems.
Photon-assisted tunneling in a resonant double barrier system was investigated
within the scattering approach. \cite{Ped98}
Various quantum dot photodetectors \cite{Lim06,Apa07,Vil07} were studied theoretically.
The photoconductivity of quantum wires and microconstrictions
with an adiabatic geometry was investigated using different calculation methods.
\cite{Gri95,Maa96,Bor97,Blo98,Gal04}
It was found that absorbtion of high-frequency electromagnetic  field
polarized in transversal direction results in
oscillations of the photoconductance as a function of the gate voltage.
Circular photogalvanic effect in bent \cite{Per05} and helical \cite{Mag03}
one-dimensional quantum wires was studied.
Radiation-induced current in quantum wires with side-coupled nanorings
was calculated. \cite{Per07}
One-dimensional quantum pump based on two harmonically oscillating
$\delta$-potentials was investigated in Refs.~\onlinecite{Bra05,Mah08}.
Various interesting photon and phonon depended effects
were predicted in carbon nanotube devices. \cite{Kib05,Kib95,Kib01,Ste04}
A number of papers are devoted to the theoretical study of
the spin-depended photogalvanic effects. \cite{Gan02,Gan07,Tar05,TI05,Gla05,Ivc04,Fed05,Ivc02}

A consistent quantitative theory of
photogalvanic effect in bulk samples was given
by Belinicher and Sturman. \cite{Bel80}
The necessary condition for appearance of the photocurrent
is the absence of the inversion symmetry in the system.
In the macroscopic samples,
the absence may be stipulated by the asymmetry
of the lattice \cite{Web08,Bel80}
or by oriented asymmetric scatterers. \cite{Ent06,Che08}

Ballistic transport regime in quasi one-dimensional nanostructueres
allows another mechanism.
The symmetry may be broken by an asymmetrically located single scatterer,
for instance, a potential barrier \cite{Tag99} or an impurity.
In the present paper, we consider one of the simplest
nanoscale system that allows generation of the direct current
induced by electromagnetical radiation.

\section{Hamiltonian}

\begin{figure}[htb]
\includegraphics[width=0.95\linewidth]{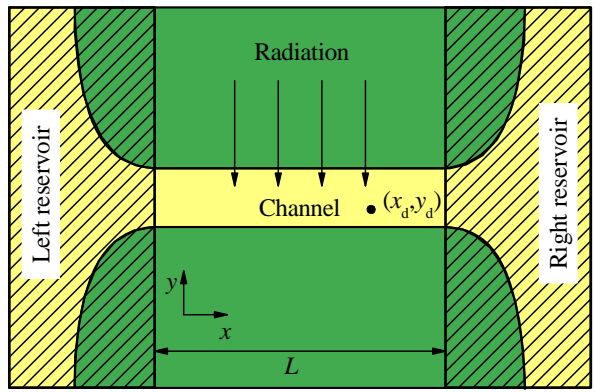}
\caption{\label{f-scheme}
(Color online) The scheme of the device. Two-dimensional
electron gas is shown by the yellow (light) area, and opaque diaphragm is
shown by the hatched area.}
\end{figure}

The purpose of the present paper
is the theoretical investigation of the photocurrent in a quantum channel
containing a single point defect.
It should be mentioned that the elastic scattering in  similar systems
has been widely discussed in literature. \cite{Yac90,Lev91,Lev92,Gur93,Lub93,Suk94,Gei93}
We consider the channel that is formed in the two-dimensional electron gas by
parabolic confining potential.
The schematic view of the device is shown in Fig.~\ref{f-scheme}.

The electron motion in the channel is described by the Hamiltonian
\begin{equation}
\hat H_0=\frac{\hat p_x^2}{2m}+\frac{\hat p_y^2}{2m}
+\frac{m\omega_0^2 y^2}{2},
\end{equation}
where $m$ is  effective electron mass, $\omega_0$ is the
frequency of parabolic confining potential, $\hat p_x$ and $\hat
p_y$ are projections of the momentum (the $x$ direction
corresponds to the axis of the channel). Eigenvalues $E_{n,p}$ and
eigenfunctions $\phi_{n,p}$ of the Hamiltonian $\hat H_0$ are
well-known
\begin{equation}
E_{n,p}=\frac{p^2}{2m}+E_n,\quad
\phi_{n,p}(x,y)=\Phi_n(y)e^{ipx/\hbar},
\end{equation}
where $E_n=\hbar\omega_0(n+1/2)$, and  $\Phi_n(y)$ are oscillator functions.
We note that the effective channel width is determined
by the number of occupied subbands and by the characteristic
oscillator length $a=\sqrt{\hbar/m \omega_0}$.

The short-range defect is modelled by the zero-range potential,
\cite{Dem88,Alb88,Bru03,Gey03,Gei93} and therefore the Hamiltonian $\hat H_{\rm d}$
of the channel with the defect is a point perturbation of the
Hamiltonian $\hat H_0$. Actually, the point perturbation is
determined by boundary conditions for the wave function at the
defect point. The similar method has been
used earlier \cite{Gey03,MP05,MP07} for modelling of point contacts.

Boundary values for the wave function are determined with the help
of the zero-range potential theory. \cite{Dem88,Alb88,Bru03,Gey03}
The theory shows that the electron wave function
$\psi(\mathbf{r})$ has the logarithmic singularity
in a vicinity of the defect point $\mathbf{r}_\mathrm{d}$.
As follows form the theory of zero-range potentials
the wave function $\psi(\mathbf{r})$ of the Hamiltonian
with the point perturbation may be represented in terms
of the Green's function. That is why the wave function has the
same singularity as the Green's function in the vicinity the point of perturbation.
We note that the form of the singularity is determined by the dimension of space only.
It is independent on energy and form of smooth confining potential.
In particular, the singularity is logarithmic in the two-dimensional case
\begin{equation}
\label{asymp}
\psi(\mathbf{r})=-u \ln |\mathbf{r}-\mathbf{r}_\mathrm{d}|+v+
R(\mathbf{r}),
\end{equation}
where $u$ and $v$ are complex coefficients, and  $R(\mathbf{r})$ is the remainder term
that tends to zero in the limit  $\mathbf{r} \to \mathbf{r}_\mathrm{d}$.
Coefficients $u$ and $v$ play the role of boundary values for the wave
function $\psi$. They are independent of $\mathbf{r}$.
The boundary conditions at the point of contact
are some linear relations between $u$ and $v$:
\begin{equation}
\label{bound}
v=P_0 u.
\end{equation}
Here the coefficient $P_0$ determines the strength of the
zero-range potential at the point $\mathbf{r}_\mathrm{d}$.
It should be noted that the zero-range potential is attractive
and the limit $|P_0|\to \infty$ corresponds to the absence of the point perturbation.

We suppose that the channel is exposed by an
external electromagnetic wave propagating in the $z$ direction and
polarized in the $y$ direction.
We assume that electron-photon interaction takes place
in the region of the channel only that may be realized
by an antenna \cite{Wyss93} or by an opaque diaphragm (Fig.~\ref{f-scheme}).
In the view of these assumptions the influence of the electromagnetic field
on the electron is described by the operator
\begin{eqnarray}
\nonumber
\hat V(t)&=&\frac{e\varepsilon \hat p_y}{m \omega}
[\theta(x)-\theta(x-L)]\cos(\omega t)\\
&=&\hat V_0 (e^{i\omega t}+e^{-i\omega t}),
\end{eqnarray}
where $\varepsilon$ is the amplitude of the electric field,
$\omega$ is the frequency of radiation,
and $\theta(x)$ is the Heaviside step function,
i.e., $\theta(x)=1$ for $x>0$ and $\theta(x)=0$ otherwise.

To obtain the electric current in the
channel we use the generalization \cite{Dat92,Maa96,Per05,Per07,Maa97}
of the Landauer--B\"uttiker formula \cite{Lan57,But86}
that takes into account the radiation
\begin{eqnarray}
\label{Initial}
\nonumber
I&=&\frac{e}{\pi\hbar}\sum\limits_{nn'l}\int\limits_0^{\infty}
[T_{n'n}^{RL}(E+l\hbar\omega,E)f_L(E)\\
&&-T_{n'n}^{LR}(E+l\hbar\omega,E)f_R(E)]dE.
\end{eqnarray}
Here $T_{n'n}^{RL}(E+l\hbar\omega,E)$  is the transmission
probability between the state with the energy $E$  and the quantum
number $n$  in the left reservoir and
the state with the energy $E+l\hbar\omega$ and the
quantum number $n'$ in the right reservoir, $f_L$ and $f_R$ are Fermi
distribution functions for the left and the right reservoirs
respectively, and $l$ is the number of absorbed photons
(negative $l$ corresponds to emission of photons).
Function $f_j$ (here $j$ means left (L) or right (R) electron reservoir) has the form
\begin{equation}
\label{Fermi}
f_j(E)=\frac{1}{\exp[(E-\mu_j)/T]+1},
\end{equation}
where $\mu_j$ is the chemical potential in the $j$-th
reservoir and $T$ is temperature.
In the paper, we consider the first order of the perturbation theory and
restrict ourselves to $l=-1,0,1$.

Transmission coefficients $T_{n'n}^{ij}(E+l\hbar\omega,E)$ can be represented
via transmission amplitudes $t_{n'n}^{ij}(E+l\hbar\omega,E)$
(here indexes $i$ and $j$ mean left ($L$) or right ($R$) reservoir)
\begin{equation}
T_{n'n}^{ij}(E+l\hbar\omega,E)=\frac{k_{n'}^l}{k_{n}^0}
\left|t_{n'n}^{ij}(E+l\hbar\omega,E)\right|^2,
\end{equation}
where wave number $k_{n}^l$ is given by
\begin{equation}
k_{n}^l=\sqrt{2m(E-E_n+l\hbar\omega)}/\hbar.
\end{equation}

\section{Photocurrent}
In the linear response approximation
the current is represented in the form
\begin{equation}
\label{Iwhole}
I=I_{\mathrm{ph}}+\mathsf{G} U_{\mathrm{bias}},
\end{equation}
where $I_{\mathrm{ph}}$ is the zero-bias photocurrent,
$\mathsf{G}$ is conductance of the system,
and $U_{\mathrm{bias}}$ is the bias voltage.
The photocurrent $I_{\mathrm{ph}}$ is given by
\begin{equation}
\label{Iph}
I_{\mathrm{ph}}=\frac{e}{\pi\hbar}\int\limits_0^\infty
f(E)\Delta T(E)dE,
\end{equation}
where
\begin{eqnarray}
\label{DT}
\nonumber
\Delta T(E)&=&\sum\limits_{l=\pm 1} \sum\limits_{nn'}
\left[T_{n'n}^{RL}(E+l\hbar\omega,E)\right.\\
&&-\left.T_{n'n}^{LR}(E+l\hbar\omega,E)\right].
\end{eqnarray}
We note that the terms with $l=0$ are cancelled in Eq.~(\ref{DT})
due to the time-reversal symmetry of elastic scattering.

To obtain the transmission probabilities
we use the concept of quasi-energy states. \cite{Zel73} Since the
Hamiltonian of the system varies periodically with time the
electron wave function may be represented in the form
\begin{equation}
\Psi(\mathbf{r},t)=\sum\limits_l \psi_l(\mathbf{r})\exp[-i(F+l\hbar\omega)t/\hbar],
\end{equation}
where $F$ is the quasi-energy.
From the Schr\"odinger equation, we obtain the following relations for
the functions $\psi_{l}(\mathbf{r})$
\begin{equation}
\nonumber
[\hat H_{\rm d}-(F+l\hbar\omega)]\psi_l(\mathbf{r})
+\hat V_0 \psi_{l+1}(\mathbf{r})+\hat V_0 \psi_{l-1}(\mathbf{r})=0.
\end{equation}
In the first order of the perturbation theory, we
restrict ourselves to $l=-1,0,1$ and express
the functions $\psi_{\pm 1}$ in terms of the Green's
function $G_{\rm d}$ of the Hamiltonian $\hat H_{\rm d}$
\begin{equation}
\label{psi_l}
\psi_{\pm 1}(\mathbf{r})
=-\int G_{\rm d}(\mathbf{r}, \mathbf{r}'; F\pm\hbar\omega)
\hat V_0\psi_0(\mathbf{r}')d\mathbf{r}',
\end{equation}
where $\psi_0(\mathbf{r})$ is determined from the equation
\begin{equation}
{(\hat H_{\rm d}-F)\psi_0(\mathbf{r})=0}.
\end{equation}
It should be mentioned
that the Green's function $G_{\rm d}(\mathbf{r}, \mathbf{r}'; F)$
is the integral kernel of the operator $(\hat H_{\rm d}-F)^{-1}$.

We note, that at real quasi-energies
the perturbation theory is inapplicable in the vicinity
of the eigenvalues $E_n$ of the Hamiltonian $\hat H_0$ since the
Green's function has the root singularities at these energies.
This phenomenon is stipulated by infinite lifetime
of the states with zero speed.
Therefore the electrons having zero speed are influenced
by electromagnetic field for an infinitely long period of time,
and hence the transition probability for these electrons is
not small. To avoid peculiarities
in the expression for $\psi_{\pm 1}$ we introduce
the complex quasi-energy $F=E+i\Gamma$.
The imaginary part $\Gamma$ of the quasi-energy is a phenomenological parameter
that may be expressed in terms
of effective state lifetime $\tau$ via the relation $\Gamma=\hbar/\tau$.
We mention that broadening $\Gamma$ may be caused by spontaneous
transitions and inelastic scattering processes.

The zero-range potentials theory \cite{Alb88,Gey03} shows that
the eigenfunction of the  Hamiltonian $\hat H_{\rm d}$ can be represented in the form
\begin{equation}
\label{psi0}
\psi_0(\mathbf{r},E)=\phi_0(\mathbf{r},E)
-\frac{\phi_0(\mathbf{r}_\mathrm{d},E)}{Q(\mathbf{r}_\mathrm{d},E)-P}
G_0(\mathbf{r},\mathbf{r}_\mathrm{d};E),
\end{equation}
where $\phi_0$ is the eigenfunction and
$G_0(\mathbf{r},\mathbf{r}';E)$ is the Green's function of the Hamiltonian $\hat H_0$,
$P=P_0+\mathrm{const}$ is a parameter that determines
the strength of the point perturbation,
and $Q(\mathbf{r}_\mathrm{d},E)$ is the Krein's Q-function that
is the renormalized Green's function. The renormalization
is obtained by subtracting of the logarithmic singularity
at $\mathbf{r}\to\mathbf{r}_\mathrm{d}$
from $G_0(\mathbf{r}_\mathrm{d},\mathbf{r},E)$.
We note, that Krein's Q-function is defined up to additive constant.
This constant determines the connection between the value of parameter $P$ and
the strength of point perturbation.
In the present paper, we focus on effects independent
of the value of the point potential, so the value of constant
is not very important for our purposes.
Since the form of singularity of the Green's function is independent of energy
we can get Krein's Q-function by the following equation \cite{Gei93}
\begin{equation}
\label{Q}
Q(\mathbf{r}_\mathrm{d},E)=\lim\limits_{\mathbf{r}\to\mathbf{r}_\mathrm{d}}
\left[G_0(\mathbf{r}_\mathrm{d},\mathbf{r},E)
-G_0(\mathbf{r}_\mathrm{d},\mathbf{r},E_0)\right].
\end{equation}
Here $E_0$ is some fixed value of energy that is smaller than the ground state energy.
Actually, $E_0$ determines the difference between parameters
$P$ from Eq. (\ref{psi0}) and $P_0$ from Eq. (\ref{bound}).
In the present paper, we take $E_0=0$ and define
the strength of the point perturbation by the parameter $P$.

To obtain the transmission amplitudes we should take $\phi_0$
as an incident wave propagating in the $n$ mode
\begin{equation}
\label{phi_0}
\phi_0(\mathbf{r},E)=\Phi_n(y)e^{ik_n^0 x}.
\end{equation}

The Green's function $G_0$ is given by equation
\begin{equation}
\label{G_0}
G_0(\mathbf{r},\mathbf{r'}; E)=\frac{im}{\hbar^2}\sum\limits_{n=0}^{\infty}
\frac{\Phi_n(y)\Phi_n^*(y')}{k_n^0}e^{ik_n^0|x-x'|},
\end{equation}
where $\mathop{\rm Re} k_{n}^0 \geq 0$ and $\mathop{\rm Im} k_{n}^0 \geq 0$.

The Green's function $G_{\rm d}$ of the
Hamiltonian $\hat H_{\rm d}$ is represented in terms of the Green's function $G_0$  of
the Hamiltonian $\hat H_0$ using the Krein resolvent formula \cite{Alb88,Bru03}
\begin{equation}
\label{G}
G_{\rm d}(\mathbf{r},\mathbf{r}'; E)=G_0(\mathbf{r},\mathbf{r}';
E) -\frac{G_0(\mathbf{r},\mathbf{r}_\mathrm{d};E)
G_0(\mathbf{r}_\mathrm{d},\mathbf{r}';E)}{Q(\mathbf{r}_\mathrm{d},E)-P}.
\end{equation}
From the asymptotics of the functions $\psi_{\pm 1}(\mathbf{r})$
at the right edge of the channel we obtain transmission amplitudes
$t_{n'n}^{RL}$ and $t_{n'n}^{LR}$. Details of the derivation are given in the Appendix.
Then we calculate transmission coefficients and the photocurrent using Eq.~(\ref{Iph}).

To write down the equation for $t_{nn'}^{RL}$ we need some preliminary notations:
$$
A=\frac{e\varepsilon a}{\sqrt{8}\hbar\omega},
\qquad f_n^l=\frac{\Phi_n(y_\mathrm{d})}{k_n^l},
$$
and
$$
\qquad
\alpha^\pm(n,l)=-\frac{m}{\hbar^2}\frac{\Phi_n(y_\mathrm{d})\exp({\pm ik_n^l x_\mathrm{d}})}
{Q(\mathbf{r}_\mathrm{d},E+l\hbar\omega)-P}.
$$
Now we can represent the amplitude in the form
\begin{equation}
\label{tnn}
t_{n'n}^{RL}(E+l\hbar\omega,E)=Ae^{ik_{n'}^l L}[t^{(1)}+t^{(2)}+t^{(3)}+t^{(4)}],
\end{equation}
where
\begin{widetext}
\begin{eqnarray}
\label{t1}
&t^{(1)}=\frac{1}{ak_{n'}^l} J_1(k_n^0,-k_{n'}^l,L)
\left[\sqrt{n+1}\delta_{n',n+1}
-\sqrt{n}\delta_{n',n-1}\right],&\\
\label{t2}
&t^{(2)}=\frac{i\alpha^-(n',l)}{ak_{n'}^l}
\left[\sqrt{n+1}f_{n+1}^l J_2(k_n^0,k_{n+1}^l)
-\sqrt{n} f_{n-1}^l J_2(k_n^0,k_{n-1}^l)\right],&\\
\label{t3}
&t^{(3)}=\frac{i\alpha^+(n,0)}{ak_{n'}^l}
\left[\sqrt{n'}f_{n'-1}^0 J_2(-k_{n'}^l,k_{n'-1}^0)
-\sqrt{n'+1}f_{n'+1}^0 J_2(-k_{n'}^l,k_{n'+1}^0)\right],&\\
\label{t4}
&t^{(4)}=\frac{\alpha^+(n,0)\alpha^-(n,l)}{ak_{n'}^l}
\sum\limits_{m=0}^\infty f_{m}^0
\left[\sqrt{m}f_{m-1}^l J_3(k_{m-1}^l,k_{m}^0)
-\sqrt{m+1}f_{m+1}^l J_3(k_{m+1}^l,k_{m}^0)\right].&
\end{eqnarray}
Here $J_1$, $J_2$ and  $J_3$ are elementary integrals:
$$
J_1(k_1,k_2,L)=\frac{1}{a}\int\limits_0^L e^{i(k_1+k_2)x}dx,
\qquad
J_2(k_1,k_2)=\frac{1}{a}\int\limits_0^L e^{ik_1 x+ik_2|x-x_\mathrm{d}|}dx,
\qquad
J_3(k_1,k_2)=\frac{1}{a}\int\limits_0^L e^{i(k_1+k_2)|x-x_\mathrm{d}|}dx.
$$
\end{widetext}
The transmission amplitude $t_{n'n}^{LR}(E+l\hbar\omega,E)$
from the right to the left reservoir
is obtained from Eq.~(\ref{tnn})
via replacing $x_\mathrm{d}$ by $L-x_\mathrm{d}$ in integrals $J_2$ and $J_3$.
\begin{figure}[htb]
\includegraphics[width=0.95\linewidth]{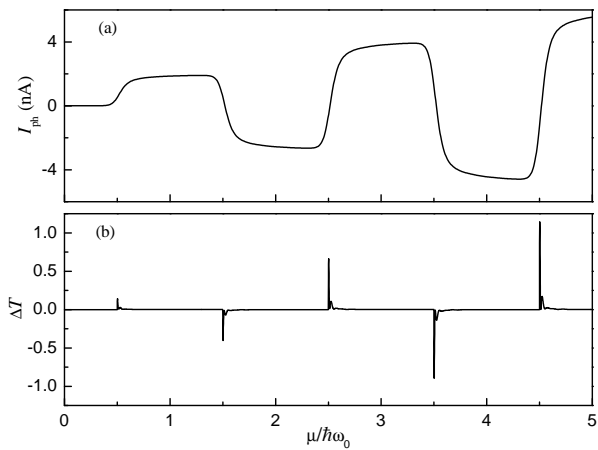}
\caption{\label{f-Iph}
(a) Dependence of the photocurrent $I_{\rm ph}$
on the chemical potential $\mu$ at
$L=0.5 \mu$m, $a=10$nm, $T=4$K,
$y_\mathrm{d}=0$, $x_\mathrm{d}=0.6L$,
$\omega=\omega_0=1.7\times10^{13}$s$^{-1}$, $\tau=6\times 10^{-11}s$,
$m=0.067 m_e$. Intensity of radiation is $0.5$W/cm$^2$.
(b) Dependence $\Delta T(\mu)$ at the same parameters.}
\end{figure}

The dependence of the photocurrent on the chemical potential $\mu$
is represented in Fig.~\ref{f-Iph}(a). The corresponding values of $\Delta T(\mu)$
are shown in Fig.~\ref{f-Iph}(b).
One can see that photocurrent oscillates with increase of the chemical potential.
The amplitude of oscillations depends linearly on the intensity of radiation.
The contribution to the photocurrent has sharp peaks in the vicinity
of the energies $E_n$. The sign of contribution is different for
even and odd values of $n$.

To explain the behavior of the photocurrent we
consider a simplified model of the system. In this model, the
channel is divided into three parts and the transmission
probabilities through each part are combined incoherently. That
means the transmission coefficients are defined by equation
\begin{equation}
T_{n'n}^{RL}=\sum_{m'm}
W_{n'm'}^{R}T_{m'm}W_{m n}^{L},
\end{equation}
where $W_{n'm'}^{R}$ and $W_{mn}^{L}$ are photon assisted transition probabilities
in the left and the right parts of the channel and
$T_{m'm}$ are elastic transmission coefficients
without taking the radiation into account.
The transition probabilities $W_{mn}^{j}$ are determined
by the same quasi-energy approach as indicated above but Eqs. (\ref{phi_0}) and
(\ref{G_0}) are used for the wave function and the Green's function
instead of Eqs. (\ref{psi0}) and (\ref{G}):
\begin{eqnarray}
\nonumber
W_{mn}^{j}&=&\frac{k_{m}^l}{k_{n}^0}
\left|\frac{A J_1(k_n^0,-k_{m}^l,L_j)e^{i k_{m}^l L_j}}{ak_{m}^l}\right|^2\\
&&\times\left[(n+1)\delta_{m,n+1}+n\delta_{m,n-1}\right].
\end{eqnarray}
Here $L_j=x_{\rm d}$ for the left part of the channel
and $L_j=L-x_{\rm d}$ for the right part.

The transmission coefficients $T_{m'm}$ are obtained from
Eq. (\ref{psi0}) and have the form
\begin{equation}
\label{Tn_dark}
T_{n'n}(E)=\delta_{n'n}-\frac{im}{\hbar^2 k_{n'}^0}
\frac{\Phi_n(y_d)\Phi_{n'}^*(y_d)}{Q(\mathbf{r}_d,E)-P}.
\end{equation}

Although the approach is sufficiently rough it
allows to obtain a simple explanation of the phenomena.
According to this approach the photocurrent is stipulated by the
difference in transmission probabilities for states with different
quantum number $n$.
Using the symmetric properties of transmission coefficient
we can represent the difference $\Delta T$ in the form
\begin{eqnarray}
\label{Tn_3PtModel}
\nonumber
\Delta T(E)&=&\sum\limits_{l=\pm 1}
\sum\limits_{m\neq n}\Delta W_{nm}(E+l\hbar\omega,E)\\
&&\times[T_{m}(E+l\hbar\omega)-T_{n}(E)],
\end{eqnarray}
where $\Delta W_{nm}=W_{nm}^L-W_{nm}^R$ and
$T_{n}(E)$ is total transmission probability for mode $n$
\begin{equation}
\label{Tn}
T_n(E)=\sum\limits_{n'}T_{n'n}(E).
\end{equation}
One can see that the electron transition contributes positively to the photocurrent
if the transmission probability increases
and contributes negatively in the opposite case.

If the scatterer is located on the axis of the channel ($y_\mathrm{d}=0$) then
the states with odd $n$ are not reflected
since the wave function vanishes at the point $\mathbf{r}_\mathrm{d}$.
Hence the electron transitions from the state with even $n$ to
the state with odd $n$ increase the transmission probability.
The number of occupied states of different
parity depends on the chemical potential. Therefore the photocurrent
oscillates with increase of the chemical potential.

Let us consider the simplest case of the
axial scatterer position ($y_\mathrm{d}=0$),
and resonance frequency of radiation ($\omega=\omega_0$).
If the defect is placed in the vicinity of the channel center
($|x_\mathrm{d}-L/2|\ll L$),
our approach gives the following asymptotic equation
for the difference of transmission coefficients:
\begin{equation}
\label{t_sim}
\Delta T(E)\simeq (-1)^N A^2
\frac{2(2N+1)L\Delta x}{(a^2 |k_N^0|)^2} e^{-2\mathop{\rm Im} k_N^0L}
\end{equation}
as $E\to E_N+0$. Here $\Delta x=x_\mathrm{d}-L/2$.
One can see from Eq.~(\ref{t_sim}) that the contribution to
the photocurrent from the vicinity of $E_n$ is positive for even
$n$ and negative for odd $n$. The photocurrent is
proportional to $\Delta x$ and it vanishes in the case of
symmetrical disposition of the defect.
The amplitude of the peak grows with increase of the mode number $N$
proportional to $2N+1$. Therefore the amplitude
of the photocurrent oscillations increases with chemical potential.
As it follows from Eq.~(\ref{t_sim}),
the contribution to the photocurrent is maximal from slow electrons.
This result is in agreement with Fig.~\ref{f-Iph}(b)
based on the more precise approach. The phenomenon is
caused by sufficiently long exposure time and large density
of states for slow electrons.

\begin{figure}[htb]
\includegraphics[width=0.95\linewidth]{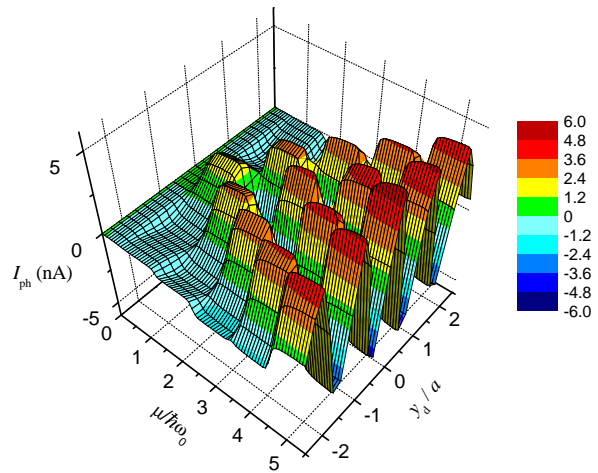}
\caption{\label{f-yd}
(Color online) Dependence of the photocurrent on the chemical potential
and the position of the scatterer.
All parameters besides $y_{\rm d}$ are the same as in Fig.~\ref{f-Iph}.}
\end{figure}

Let us discuss the effect of the scatterer position.
The dependence of the photocurrent on the chemical potential
and the scatterer position is represented in Fig.~\ref{f-yd}.
One can see that
the sequence of peaks and dips of the photocurrent is changed
if the scatterer is shifted from the channel axis ($y_\mathrm{d}\neq0$).
The photocurrent tends to zero when the defect is placed sufficiently far from
the channel axis ($|y_\mathrm{d}|\gg a$) because
the defect does not change the transmission probability in this case.

\begin{figure}[htb]
\includegraphics[width=0.95\linewidth]{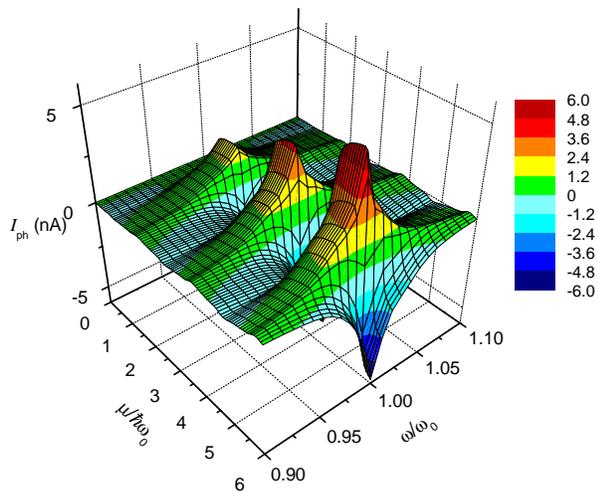}
\caption{\label{f-omega}
(Color online)
Dependence of the photocurrent on the chemical potential
and the frequency of the radiation.
All parameters besides $\omega$ are the same as in Fig.~\ref{f-Iph}.}
\end{figure}

The dependence of the photocurrent on the frequency of radiation
and the  chemical potential is represented in Fig.~\ref{f-omega}.
One can see that the photocurrent decreases with deviation
of the frequency from the resonance value $\omega_0$.

Our numerical analysis shows that more simplified 'incoherent'
method of calculation is in qualitatively agreement
with more precise method based on Eqs.~(\ref{tnn})-(\ref{t4}).
But asymptotics given by Eq.~(\ref{t_sim}) for  $\Delta T(E)$
is not valid for very small values of $\Gamma$.
According to Eq.~(\ref{t_sim}) $\Delta T(E)\to \infty$ at $\mu=E_N$ as $\Gamma\to 0$.
However, more precise analysis based on Eqs.~(\ref{tnn})-(\ref{t4})
shows that $\Delta T(E)$ remains finite.
The difference between the approaches is based on taking the quantum coherence
into account. We mention, that all figures are plotted using Eqs.~(\ref{tnn})-(\ref{t4}).

\section{Photoconductance}
Let us consider the effect of the radiation
on the conductive properties of the device.
According to Eqs. (\ref{Initial}) and (\ref{Iwhole})
the conductance $\mathsf{G}$ can be represented in the form
\begin{equation}
\label{conductance}
\mathsf{G}(\mu,T)=
\mathsf{G}_0\int\limits_0^\infty \left(-\frac{\partial f}{\partial E}\right)
\overline{T}(E) dE,
\end{equation}
where $\mathsf{G}_0$ is the conductance quantum and $\overline{T}(E)$ is given by
\begin{eqnarray}
\nonumber
\overline{T}(E)&=&
\frac{1}{2}\sum\limits_{l=-1}^{1} \sum\limits_{nn'}
\left[T_{n'n}^{RL}(E+l\hbar\omega,E)\right.\\
&&\left.+T_{n'n}^{LR}(E+l\hbar\omega,E)\right].
\end{eqnarray}

It should be noted that $\overline{T}(E)$ contains contributions from elastic
and inelastic processes for $l=0$ and $l\neq 0$ respectively.
Hence to obtain the photoconductance
we have to find the modification of the transmission coefficients
$T_{n'n}^{ij}(E,E)$ that corresponds to electron motion
without absorbtion or emission of photons.
These coefficients could not be found from the first-order perturbation theory
since the corrections for the transmission amplitudes
should be the second-order terms.
Therefore, to find the corrections we use the current conservation law
\begin{equation}
\label{TR}
\sum\limits_{l=-1}^1\sum\limits_{n'}
\left[T_{n'n}^{RL}(E+l\hbar\omega,E)+R_{n'n}^{RL}(E+l\hbar\omega,E)\right]=1,
\end{equation}
where $R_{n'n}^{RL}(E+l\hbar\omega,E)$ are reflection coefficients.
Similar relation is valid for electrons travelling
from the right to the left reservoir.
We mention that the coefficients $T_{n'n}^{ij}(E,E)$ were cancelled in
Eq.~(\ref{Iph}) due to the time reversal symmetry.

Using Eq.~(\ref{TR}), we represent $\mathsf{G}(\mu,T)$ in the form
\begin{equation}
\mathsf{G}(\mu,T)=\mathsf{G}_{\mathrm{dark}}(\mu,T)+\mathsf{G}_{\mathrm{ph}}(\mu,T),
\end{equation}
where $\mathsf{G}_{\mathrm{dark}}$ is 'dark' quasiballistic conductance of the channel
with the defect and $\mathsf{G}_{\mathrm{ph}}$ is photoconductance.

At the zero temperature the conductance $\mathsf{G}_{\mathrm{dark}}$
may be represented in the form
\begin{equation}
\mathsf{G}_{\mathrm{dark}}(\mu)=\mathsf{G}_0\sum\limits_{n'}T_{n'n}(\mu),
\end{equation}
where $T_{n'n}$ are transmission coefficients for elastic scattering
given by Eq.~(\ref{Tn_dark}).

\begin{figure}[htb]
\includegraphics[width=0.95\linewidth]{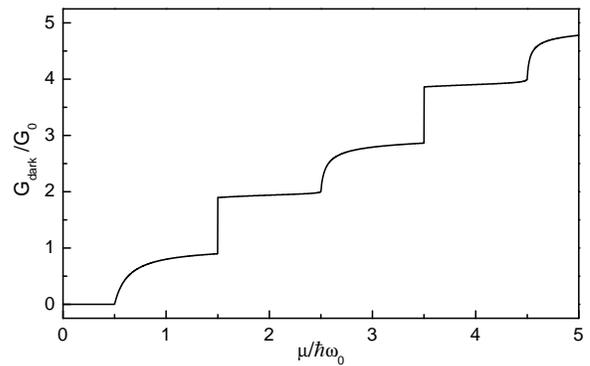}
\caption{\label{f-Gd}
Dependence of the 'dark' conductance $\mathsf{G}_{\mathrm{dark}}$
on the chemical potential $\mu$ at $y_{\rm d}=0$ and $T=0$.}
\end{figure}

Using Eqs. (\ref{G_0}) and (\ref{Tn_dark}), we can represent
the zero-temperature conductance $\mathsf{G}_{\mathrm{dark}}(\mu)$ in the form
\begin{equation}
\mathsf{G}_{\mathrm{dark}}(\mu)=
\mathsf{G}_0\left( N(\mu)-
\frac{[\mathop{\rm Im} Q(\mathbf{r}_d,\mu)]^2}{|Q(\mathbf{r}_d,\mu)-P|^2}\right).
\end{equation}
Here
$N(E)=[E/\hbar\omega-1/2]$ is the number of occupied subbands
($[x]$ means integer part of $x$).
It is clear, that at finite temperatures the conductance
$\mathsf{G}_{\mathrm{dark}}(\mu,T)$ is given
by the integral similar to Eq.~(\ref{conductance}).
The dependence $\mathsf{G}_{\mathrm{dark}}(\mu)$
in the case of axial scatterer position ($y_{\rm d}=0$)
is shown in Fig.~\ref{f-Gd}.
One can see that the steps corresponding to even values of $n$ are smoothed
due to scattering on the defect while the steps with odd values of $n$ are conserved
because electrons with odd values of $n$ are not scattered.

Photoconductance $\mathsf{G}_{\mathrm{ph}}$ is given by equation
\begin{equation}
\label{Gph}
\mathsf{G}_{\rm ph}=\mathsf{G}_0\int\limits_0^\infty \left(-\frac{\partial f}{\partial E}\right)
T^{\rm ph}(E)dE,
\end{equation}
where
\begin{eqnarray}
\label{Tav2}
 T^{\rm ph}(E)=\sum\limits_{n}
\left[\widetilde{T}_{n}(E)R_{n}(E)
-\widetilde{R}_{n}(E)T_{n}(E)\right].
\end{eqnarray}
Here  $T_{n}(E)$ and $R_{n}(E)$ are total elastic
transmission and reflection probabilities
for mode $n$ while $\widetilde{T}_{n}(E)$ and $\widetilde{R}_{n}(E)$
are total photon-assisted transmission and reflection probabilities
for mode $n$.
Probability $T_n$ is given by Eq.~(\ref{Tn}) and $R_n$ is defined by
$R_n(E)=1-T_n(E)$.
According to Eq. (\ref{TR}), $\widetilde{T}_{n}(E)$ and $\widetilde{R}_{n}(E)$
are given by the following equations:
\begin{eqnarray}
\nonumber
\widetilde{T}_{n}(E)&=&\frac{1}{2}\sum\limits_{l=\pm 1}\sum\limits_{n'}
\left[T_{n'n}^{RL}(E+l\hbar\omega,E)\right.\\
&&\left.+T_{n'n}^{LR}(E+l\hbar\omega,E)\right]
\end{eqnarray}
and
\begin{eqnarray}
\nonumber
\widetilde{R}_{n}(E)&=&\frac{1}{2}\sum\limits_{l=\pm 1}\sum\limits_{n'}
\left[R_{n'n}^{LL}(E+l\hbar\omega,E)\right.\\
&&\left.+R_{n'n}^{RR}(E+l\hbar\omega,E)\right].
\end{eqnarray}
\begin{figure}[htb]
\includegraphics[width=0.95\linewidth]{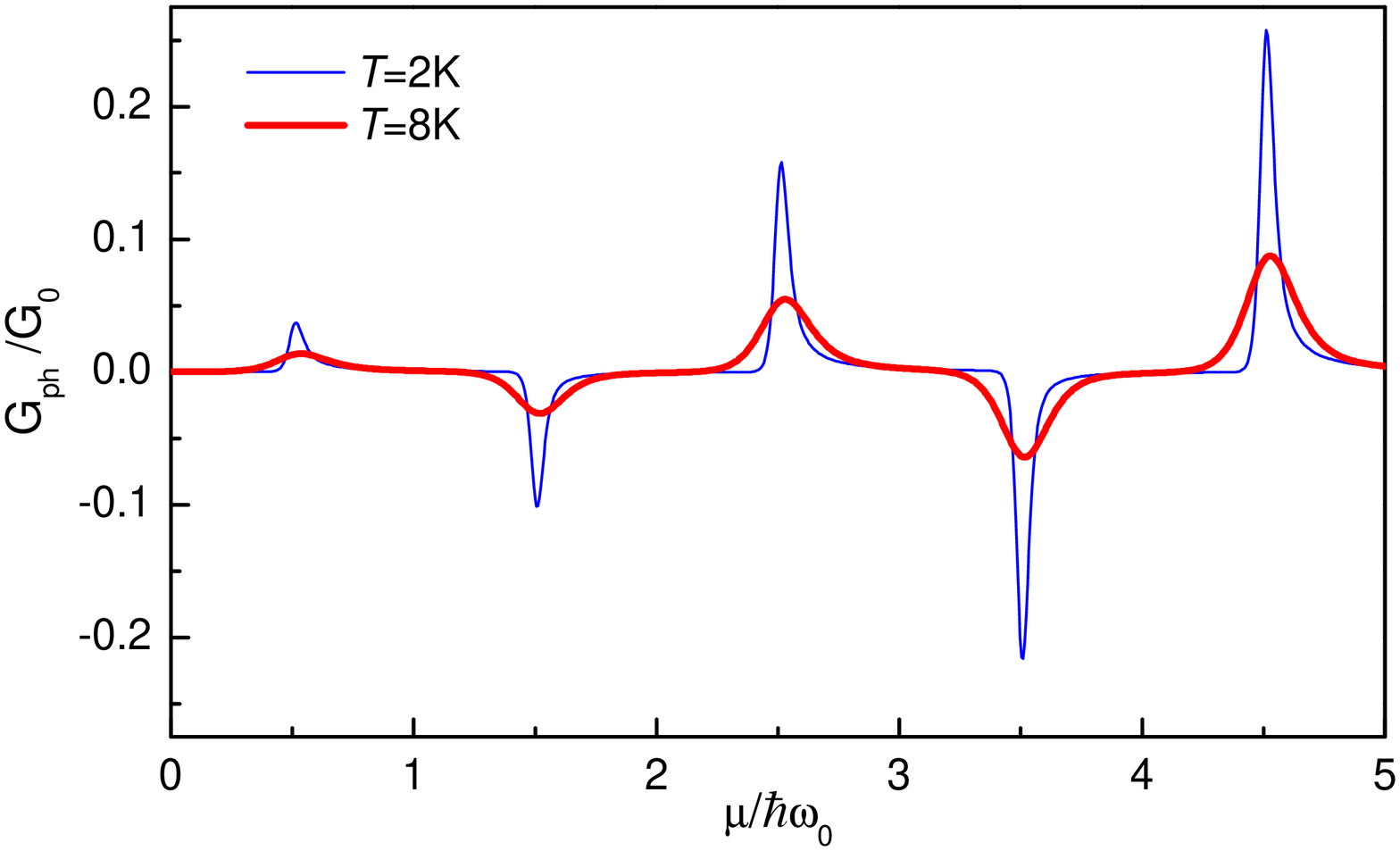}
\caption{\label{f-Gph}
(Color online) Dependence of the photoconductance
$\mathsf{G}_{\mathrm{ph}}$ on the chemical potential $\mu$ at $y_{\rm d}=0$.
Thin blue line:  $T=2K$, thick red line: $T=8K$.
Other parameters are the same as in Fig.~\ref{f-Iph}.}
\end{figure}

The photoconductance as a function of the chemical potential
oscillates as well as the photocurrent.
The dependence $\mathsf{G}_{\mathrm{ph}}(\mu)$
is shown  in Fig.~\ref{f-Gph}.
One can see from Eq. (\ref{Gph}) that the dependence $\mathsf{G}_{\mathrm{ph}}(\mu)$
contains the derivative of the Fermi function
in contrast to the dependence $I_{\mathrm{ph}}(\mu)$,
therefore peaks and dips of the photoconductance
are sharper and more sensitive to the growth
of the temperature. The increase of temperature leads
to smoothing of peaks and decrease of their amplitudes.
The amplitude of the photoconductance
peaks is proportional to intensity of radiation.
If the defect is placed in the central cross-section of the channel
then the photoconductance is proportional to squared length of the channel.

\section{Conclusion}

Photon-induced electron transport in the quantum
channel with a single point defect is investigated
using modified Landauer--B\"uttiker formalism.
The dependence of the photocurrent on the electron
chemical potential is studied both analytically and numerically.
We have shown that the photocurrent
is stipulated by different transmission probabilities for different
electron subbands in the channel.
If the defect is placed on the axis of the channel then
the odd electron subbands are not reflected by the defect since corresponding
wave function vanishes at the point of the scatterer.
Thus the photon-induced transitions between
electron subbands of different parity
can either increase or decrease the transmission probability.
The probability of photon-induced transition  depends on the distance
from the channel edge to the defect.
Therefore, the direct current arises in the case of asymmetric scatterer position.
We have shown that the photocurrent oscillates as a function of chemical potential.
The direction of the current is determined by the number
of occupied subbands of different parity.
The sequence of oscillating minima and maxima is changed if the scatterer is placed
aside the channel axis.
The photocurrent is proportional to the intensity of radiation
and grows with increase of the lifetime $\tau$.
The amplitude of oscillations reaches
maximum when the frequency $\omega$ of radiation coincides
with the characteristic frequency $\omega_0$
of the confining potential in the channel.
This feature of the system gives the possibility
to vary the resonance frequency by changing the geometry of the channel.

If the scatterer is placed in the central cross-section of the channel
the photocurrent is absent but the defect
influences the conductance of the system.
The total conductance of the system may be represented as a sum of
the 'dark' quasiballistic conductance $\mathsf{G}_{\rm dark}$ and
the photoconductance $\mathsf{G}_{\rm ph}$ that is
proportional to the intensity of radiation.
The dependence of the photoconductance on the chemical potential contains
sharp peaks and dips in the vicinity of steps of the ballistic conductance.
In the case of axial defect position, the sign of the photoconductance
is determined by the parity of the highest occupied subband in the
channel at the zero temperature.
The nature of the photoconductance oscillations is similar to the nature of the
photocurrent oscillations.
However, the dependence of the photoconductance
on the chemical potential is sharper and more sensitive
to the temperature than the dependence of the photocurrent.
The grows of temperature leads to the smoothing of photoconductance peaks
and decrease of their amplitudes.

\section*{ACKNOWLEDGEMENTS}

The authors are grateful to V. A. Margulis for helpful discussions.
The work has been supported by Grant of the President of the Russian Federation for
young scientists (grant MK-4480.2007.2)
and by Russian Foundation for Basic Research (grant 08-02-01035).

\section*{APPENDIX}

To obtain the transmission amplitude $t_{nn'}^{RL}$
from the left to the right reservoir
we have to compare the asymptotics of the wave function
at the left and the right edge of the channel.
The wave function $\psi_{\pm 1}$  is given by Eq.~(\ref{psi_l}).
We take the incident wave $\phi_0$ from the left reservoir
in the form Eq.~(\ref{phi_0}).
Using Eq.~(\ref{psi0}) for $\psi_0$ and Eq.~(\ref{G}) for $G_\mathrm{d}$
we may represent the function $\psi_{\pm 1}$ as a sum of four terms
\begin{equation}
\psi_{\pm 1}(\mathbf{r})=\psi^{(1)}(\mathbf{r})+\psi^{(2)}(\mathbf{r})+
\psi^{(3)}(\mathbf{r})+\psi^{(4)}(\mathbf{r}),
\end{equation}
where functions $\psi^{(j)}(\mathbf{r})$ are given by
\begin{widetext}
\begin{equation}
\label{psi1}
\psi^{(1)}(\mathbf{r})=-\int G_{0}(\mathbf{r}, \mathbf{r}'; F\pm\hbar\omega)
\hat V_0\phi_0(\mathbf{r}',F) d\mathbf{r}',
\end{equation}
\begin{equation}
\label{psi2}
\psi^{(2)}(\mathbf{r})=\frac{G_0(\mathbf{r},\mathbf{r}_{\mathrm{d}}, F\pm\hbar\omega)}
{Q(\mathbf{r}_{\mathrm{d}}, F\pm\hbar\omega)-P}
\int G_{0}(\mathbf{r}_{\mathrm{d}}, \mathbf{r}'; F\pm\hbar\omega)
\hat V_0\phi_0(\mathbf{r}',F) d\mathbf{r}',
\end{equation}
\begin{equation}
\label{psi3}
\psi^{(3)}(\mathbf{r})=\frac{\phi_0(\mathbf{r}_{\mathrm{d}}, F)}
{Q(\mathbf{r}_{\mathrm{d}}, F)-P}
\int G_{0}(\mathbf{r}, \mathbf{r}'; F\pm\hbar\omega)
\hat V_0 G_{0}(\mathbf{r}', \mathbf{r}_{\mathrm{d}}; F) d\mathbf{r}',
\end{equation}
\begin{equation}
\label{psi4}
\psi^{(4)}(\mathbf{r})=-\frac{\phi_0(\mathbf{r}_{\mathrm{d}}, F)}
{Q(\mathbf{r}_{\mathrm{d}}, F)-P}
\frac{G_0(\mathbf{r},\mathbf{r}_{\mathrm{d}}, F\pm\hbar\omega)}
{Q(\mathbf{r}_{\mathrm{d}}, F\pm\hbar\omega)-P}
\int G_{0}(\mathbf{r}, \mathbf{r}'; F\pm\hbar\omega)
\hat V_0 G_{0}(\mathbf{r}', \mathbf{r}_{\mathrm{d}}; F) d\mathbf{r}'.
\end{equation}

Taken into consideration properties of oscillator functions $\Phi_n$ we have
\begin{equation}
\label{p_y_Phi_n} \hat p_y \Phi_n(y)=\frac{i\hbar}{a}
\left\{\sqrt{\frac{n+1}{2}}\Phi_{n+1}(y)-\sqrt{\frac{n}{2}}\Phi_{n-1}(y)\right\}.
\end{equation}
Using Eqs.~(\ref{phi_0}), (\ref{G_0}), and (\ref{p_y_Phi_n}) we obtain
\begin{equation}
\label{V0phi}
\hat V_0\phi_0(\mathbf{r}')=\frac{ie\epsilon\hbar}{\sqrt{8}m\omega a}
\left\{\sqrt{n+1}\Phi_{n+1}(y')-\sqrt{n}\Phi_{n-1}(y')\right\}
\left[\theta(x')-\theta(x'-L)\right]\exp({ik_{n}^0 x'}),
\end{equation}
\begin{equation}
\label{V0G}
\hat V_0 G_{0}(\mathbf{r}', \mathbf{r}_{\mathrm{d}};F)=
\frac{e\epsilon}{\sqrt{8}\hbar\omega a}
\left[\theta(x')-\theta(x'-L)\right]
\sum\limits_{n'=0}^{\infty}
\left\{\sqrt{n}\Phi_{n-1}(y')-\sqrt{n+1}\Phi_{n+1}(y')\right\}
\frac{\exp(ik_{n'}^0 |x'-x_\mathrm{d}|)}{ak_{n'}^{0}}.
\end{equation}
Now we substitute Eqs. (\ref{V0phi}) and (\ref{V0G})
into Eqs. (\ref{psi1})--(\ref{psi4}) and perform the integration over $\mathbf{r}'$.
One can see that integration over $x'$ includes only finite interval
due to the $\theta$-functions.
Integration over $y'$ may be performed easily
due to the orthogonality of the oscillator functions $\Phi_n(y)$.
For example, integration in equation (\ref{psi1}) gives for $x \geq L$
\begin{equation}
\psi^{(1)}(\mathbf{r})=\frac{e\epsilon}{\sqrt{8}\hbar\omega a}
\sum\limits_{n'=0}^\infty
\left\{\sqrt{n+1}\delta_{n',n+1}-\sqrt{n}\delta_{n',n-1}\right\}
\frac {\Phi_{n'}(y)}{a k_{n'}^l} \exp (ik_{n'}^l x)
\int\limits_{0}^L dx' \exp [i(k_{n'}^l-k_n^0)x'].
\end{equation}
\end{widetext}
Then we obtain the following form for
functions $\psi^{(j)}$ at the right edge of the channel ($x \geq L$)
\begin{equation}
\psi^{(j)}(\mathbf{r})=\frac{e\epsilon a}{\sqrt{8}\hbar\omega}
\sum\limits_{n'=0}^{\infty}t_{n'n}^{(j)}\Phi_{n'}(y)\exp({ik_{n'}^l x}),
\end{equation}
where coefficients $t_{n'n}^{(j)}$ are given by Eqs.~(\ref{t1})--(\ref{t4}).
It should be mentioned that we have to take into account
the exponential factor $\exp(ik_n^l L)$ in the transmission
amplitude since we deal with complex
quasienergies and consequently complex wave numbers $k$.
This factor does not vanish during the current calculation hence it
should be conserved in the equation for transmission amplitude.

To get the transmission amplitude $t_{nn'}^{LR}$ from the right to the left reservoir
we need do consider incident wave that is injected from the right
edge of the channel and propagate to the left edge.
One can see that the computation remains the same if we invert the direction of
the $x$ axis. Therefore the final expressions for transmission amplitude
may be obtained from Eqs.~(\ref{t1})--(\ref{t4}) by replacement of $x_{\rm d}$
with $L-x_{\rm d}$.

\end{document}